\overfullrule=0pt
\input harvmac
\def\a{{\alpha}}

\def\l{{\lambda}}
\def\lh{{\widehat\lambda}}
\def\b{{\beta}}

\def\g{{\gamma}}

\def\d{{\delta}}

\def\s{{\sigma}}
\def\r{{\rho}}
\def\s{{\sigma}}

\def\N{{\nabla}}
\def\Nb{{\overline\nabla}}

\def\O{{\Omega}}
\def\Ob{{\overline\O}}

\def\o{{\omega}}
\def\oh{{\widehat\omega}}

\def\p{{\partial}}
\def\pb{{\overline\partial}}
\def\t{{\theta}}

\def\oh{{\widehat\o}}
\def\L{{\Lambda}}
\def\Lt{{\widetilde\Lambda}}

\def\Pib{{\overline\Pi}}
\def\Jb{{\overline J}}
\def\AA{{\cal A}}
\def\BB{{\cal B}}
\def\GG{{\cal G}}
\def\HH{{\cal H}}
\def\Qt{\widetilde Q}

\baselineskip12pt

\Title{ \vbox{\baselineskip12pt
}}
{\vbox{\centerline
{The Non-minimal Heterotic Pure Spinor String}
\bigskip
\centerline{ in a Curved Background  }
}}
\smallskip
\centerline{Osvaldo Chandia\foot{e-mail: ochandiaq@gmail.com}, }
\smallskip
\centerline{\it Facultad de Artes Liberales, Universidad Adolfo Ib\'a\~nez}
\centerline{\it Facultad de Ingenier\'{\i}a y Ciencias, Universidad Adolfo Ib\'a\~nez}
\centerline{\it Diagonal Las Torres 2640, Pe\~nalol\'en, Santiago, Chile}

\bigskip

\noindent
We study the non-minimal pure spinor string in a curved background. We find that the minimal BRST invariance implies the existence of a non-trivial stress-energy tensor for the minimal and non-minimal variables in the heterotic curved background. We find constraint equations for the $b$ ghost. We construct the $b$ ghost as a solution of these constraints.

\Date{November 2013}


\lref\BerkovitsFE{
  N.~Berkovits,
  ``Super Poincare covariant quantization of the superstring,''
JHEP {\bf 0004}, 018 (2000).
[hep-th/0001035].
}

\lref\BerkovitsNN{
  N.~Berkovits,
  ``Cohomology in the pure spinor formalism for the superstring,''
JHEP {\bf 0009}, 046 (2000).
[hep-th/0006003].
}

\lref\BerkovitsQX{
  N.~Berkovits and O.~Chandia,
  ``Massive superstring vertex operator in D = 10 superspace,''
JHEP {\bf 0208}, 040 (2002).
[hep-th/0204121].
}

\lref\BedoyaNP{
  O.~A.~Bedoya and N.~Berkovits,
  ``GGI Lectures on the Pure Spinor Formalism of the Superstring,''
[arXiv:0910.2254 [hep-th]].
}

\lref\BerkovitsBT{
  N.~Berkovits,
  ``Pure spinor formalism as an N=2 topological string,''
JHEP {\bf 0510}, 089 (2005).
[hep-th/0509120].}

\lref\ChandiaIX{
  O.~Chandia,
  ``The b Ghost of the Pure Spinor Formalism is Nilpotent,''
Phys.\ Lett.\ B {\bf 695}, 312 (2011).
[arXiv:1008.1778 [hep-th]].
}

\lref\JusinskasYCA{
  R.~Lipinski Jusinskas,
  ``Nilpotency of the b ghost in the non-minimal pure spinor formalism,''
JHEP {\bf 1305}, 048 (2013).
[arXiv:1303.3966 [hep-th]].
}

\lref\JusinskasSHA{
  R.~L.~Jusinskas,
  ``Notes on the pure spinor b ghost,''
JHEP {\bf 1307}, 142 (2013).
[arXiv:1306.1963 [hep-th]].
}

\lref\BakhmatovFPA{
  I.~Bakhmatov and N.~Berkovits,
  ``Pure Spinor b-ghost in a Super-Maxwell Background,''
[arXiv:1310.3379 [hep-th]].
}

\lref\BerkovitsUE{
  N.~Berkovits and P.~S.~Howe,
  ``Ten-dimensional supergravity constraints from the pure spinor formalism for the superstring,''
Nucl.\ Phys.\ B {\bf 635}, 75 (2002).
[hep-th/0112160].
}

\lref\ChandiaHN{
  O.~Chandia and B.~C.~Vallilo,
 ``Conformal invariance of the pure spinor superstring in a curved background,''
JHEP {\bf 0404}, 041 (2004).
[hep-th/0401226].
}

\lref\ChandiaVP{
  O.~Chandia and M.~Tonin,
  ``BRST anomaly and superspace constraints of the pure spinor heterotic string in a curved background,''
JHEP {\bf 0709}, 016 (2007).
[arXiv:0707.0654 [hep-th]].
}

\lref\ChandiaXI{
  O.~Chandia,
  ``A Note on the classical BRST symmetry of the pure spinor string in a curved background,''
JHEP {\bf 0607}, 019 (2006).
[hep-th/0604115].
}

\newsec{Introduction}

The pure spinor formalism of the superstring was constructed more than a decade ago \BerkovitsFE. The idea is to add a constrained ghost, which satisfies the pure spinor condition. The string sigma model is constructed in a way that it is conformal invariant. Berkovits noted that the conformal invariance of the model was not enough to get the superstring physical spectrum and invented a nilpotent charge with the help of a pure spinor variable. It turns out that the cohomology of this nilpotent charge gives the physical superstring spectrum and nothing else \BerkovitsNN. Unlike RNS, the pure spinor formalism does not need to make a projection to get the physical spectrum and space-time supersymmetry is manifest. In fact, not only massless sates are described in terms of superfields, massive states can also be described in this language \BerkovitsQX. Many applications of the formalism, like computing manifestly supersymmetric scattering amplitudes, were developed later (see the review \BedoyaNP). 

Despite the success in reproducing known results in other formalisms and obtaining new results, the pure spinor formalism is not understood completely. Perhaps, the most important lacking ingredient is a symmetry of the world-sheet action that allows quantization. In other words, it is not known what is the fixed gauge symmetry that implies the existence of the pure spinor BRST charge. Instead of facing directly this problem, one could continue the program and determine some features that would lead, eventually, to solve the previous issue. One of this features is the inclusion of a $b$ ghost. Since the string model does not require a pair of $(b, c)$ ghosts, the BRST ghosts of the parameterization invariance of the string world-sheet, they have to be constructed as a functions of the string model variables. However, the minimal pure spinor formalism of \BerkovitsFE\ is not suitable to define a $b$ ghost. Berkovits introduced new variables and constructed a non-minimal pure spinor formalism \BerkovitsBT. The cohomology of the modified BRST charge does not change respect to the minimal version of the pure spinor formalism  \BerkovitsBT.  The nilpotency of the $b$ ghost was verified in \ChandiaIX\ and \JusinskasYCA. The conjugate $c$ ghost was constructed in \JusinskasSHA. All this was done in flat space-time background. Recently, the construction of the $b$ ghost in a super Maxwell background was done in \BakhmatovFPA. The natural question is to determine the $b$ ghost in a generic curved background. This is the purpose of this paper.

The idea is to determine a string world-sheet action in a heterotic curved background that it os consistent with the pure spinor BRST symmetry. We find that the stress-energy tensor has the form
\eqn\gent{ T = T_0 + T_1 ,}
where $T_0$ is the stress-energy tensor for the minimal variable and $T_1$ is an expression that it is reduced to the correct limit on a flat space-time background. The expression for $T$ is determined in section 3. Once $T$ is determined, the $b$ ghost is obtained through the relation $Qb = T$. The result has the form
\eqn\gen{Êb = b_0 + f (\O_\a) ,}
where $b_0$ has the same dependence on world-sheet than the ghost in flat background space-time and $f$ is conformal weight two which depends linearly on the scalar part of the Lorentz connection $\O_\a$. Note that $\O_\a = {1\over 4} \N_\a \Phi$, where $\Phi$ is the dilaton superfield. Then, in the flat space-time background $\Phi$ vanishes and the $b$ ghost has the correct limit. Note also that there are backgrounds where the dilaton superfield is constant, then it would be possible that in such cases, the $b$ ghost has the same dependence on world-sheet fields thant the $b$ ghost in flat space-time background.  

The plan of the paper is as it follows. In section 2, the non-minimal pure spinor formalism in flat background is reviewed. In section 3, the minimal pure spinor formalism in the heterotic curved background is reviewed. This system was studied in \BerkovitsUE, where it was shown that nilpotency of the BRST charge implies that the background satisfies the ten-dimensional supergravity equations of motion and the $N=1$ super Yang-Mills equations of motion in a curved background. Note that this background was shown to be conformal invariant at one-loop  \ChandiaHN\ and the one-loop BRST anomaly was studied in \ChandiaVP. The BRST transformations for the world-sheet fields in a curved heterotic background were determined in \ChandiaXI.

The BRST transformation of the minimal and non-minimal variables are obtained in the section 4. They are obtained as consequence of the trivial cohomology of the non-minimal contribution to the BRST charge. This fact was noted in \BerkovitsBT\ in flat space-time background. We generalize this fact to the heterotic curved background. In section 5, we determine the world-sheet  action of the non-minimal pure spinor string in the heterotic curved background. Here, the stress-energy tensor receives a non-trivial contribution from the non-minimal sector. It is important to determine the stress-energy tensor because it will allow to find constraint equations for the $b$ ghost. This is done in section 6.  The constraints equations come from the definition for the $b$ ghost. It satisfies $Qb=T$, where $Q$ is the BRST charge and $T$ is the stress-energy tensor. Finally, in section 7, we solve the constraint equations for the $b$ ghost and determine that the resulting construction has the correct flat space-time background limit. 

\newsec{The Non-minimal Pure Spinor String on a Flat Background}

In this section we review the non-minimal pure spinor formalism \BerkovitsBT\ on a flat background. The action is given by
\eqn\action{ S = S_0 + \int d^2z ~ \oh^\a\pb\lh_\a + s^\a \pb r_\a ,}
where $S_0$ is the minimal action which is given by
\eqn\actionp{ S_0  = \int d^2z ~ \ha ~ \p X^m \pb X_m + p_\a \pb \t^\a + \o^\a \pb \l^\a ,}
where $(X^m, \t^\a)$ are the coordinates of $N=1$ ten-dimensional superspace, $p_\a$ is the canonical conjugate of $\t^\a$. The minimal $\l$ and the non-minimal $(\lh, r)$ ghosts are constrained to satisfy 
\eqn\psc{ \l\g^m\l=\lh\g^m\lh=\lh\g^m r = 0 ,}
where $\g_m^{\a\b}$ and $\g^m_{\a\b}$ are the symmetric gamma matrices in ten dimensions.

In order to preserve these constraints, the canonical conjugate ghosts $\o, \oh$ and $s$ are defined up to the gauge transformations
\eqn\gaugegh{ \eqalign{ &\d\o_\a=(\l\g_m)_\a \L^m,\quad \d s^\a = (\g^m\lh)^\a \Lt_m ,\cr
&\d\oh^\a=(\g^m\lh)^\a \Lt_m - (\g^m r)^\a \Lt_m .\cr}}

The quantization of this system is performed after the inclusion of a nilpotent charge which is identified with a BRST charge. It is given by
\eqn\brst{ Q = Q_0 + Q_1 ,}
where
\eqn\quis{ Q_0 = \oint dz ~\l^\a d_\a,\quad Q_1 = \oint dz ~ \oh^\a r_\a ,}
are the the BRST charges for the minimal and non-minimal pure spinor variables.  Note that the BRST charge is nilpotent because both $Q_0$ and $Q_1$ are nilpotent and anticommute. 

The cohomology of the minimal BRST operator, the first term in \brst, describes the physical superstring states. For massless states, the (unintegrated) vertex operator is $U=\l^\a A_\a(X,\t)$. This state is in the cohomology of $Q_0$ if $Q_0 U = 0$ and $U \sim U + Q_0 U$. These conditions imply that $U$ contains the photon an the photino as physical degrees of freedom and nothing else \BerkovitsFE. Similarly, vertex operators for higher mass states can be defined and cohomology conditions put the superfields on-shell. This is the case for the first massive state where the only physical states are the massive spin-$3/2$ multiplet \BerkovitsQX. The cohomology of the non-minimal pure spinor BRST operator \brst\ is equivalent to the cohomology of the minimal pure spinor BRST operator in \brst\ because the non-minimal contribution has trivial cohomology \BerkovitsBT. Consequently, the action of the non-minimal model is BRST equivalent to the action of the minimal model. In fact
\eqn\Qequiv{ÊS = S_0 + Q \int d^2z ~ s^\a \pb \lh_\a .}
Below, this relation will be used to determine the superstring world-sheet action on a curved background.

\newsec{The Minimal Heterotic Pure Spinor String on a Curved Background}

In this section we review the construction of the action for the heterotic string in a curved background (see the appendix for a short review of our conventions). The action can be obtained by adding to the flat action of \action\ the integrated vertex operator and then covariantize respect to background invariance. The action becomes \BerkovitsUE\
\eqn\scurvmin{ \eqalign{ S_0 =  \int d^2z ~ [ \ha \Pi^a \Pib^b \eta_{ab} + & \ha \Pi^A \Pib^B B_{BA} + d_\a ( \Pib^\a + \Jb^I W_I^\a ) + \l^\a \o_\b \Jb^I U_{I\a}{}^\b \cr &+ \o_\a \Nb \l^\a + (\rho_{\cal A} \N \rho_{\cal A} ) ] +  S_{FT} ,\cr}}
where $\Pi^A$ and $\Pib^A$ are defined from the background supevielbein $E_M{}^A$ and the superfield coordinates $Z^M$ as
\eqn\pis{
\Pi^A = \p Z^M E_M{}^A,\quad \Pib^A = \pb Z^M E_M{}^A .}
The variable $d_\a$ is interpreted as the world-sheet generator for translations in superspace. The world-sheet covariant derivative on the pure spinor variable is defined by
\eqn\covder{
\Nb\l^\a = \pb\l^\a + \l^\b \Ob_\b{}^\a ,}
where $\O_\b{}^\a = \pb Z^M \O_{M\b}{}^\a$ with $\O$ being the Lorentz connection. Note that the connection $\O_{A\a}{}^\b = E_M{}^A \O_{M\a}{}^\b$ has the index structure 
\eqn\connection{ \O_{A\a}{}^\b = \O_A \d_\a^\b + {1\over4} (\g_{ab})_\a{}^\b \O_A{}^{ab} ,}
where $\O_A$ is the scalar connection and $\O_A{}^{ab}$ is the usual Lorentz connection. The right-moving heterotic fermions $\rho_{\cal A}$ transform in the fundamental representation of $E_8 \times E_8$ or $SO(32)$ and its covariant derivative is defined such that
$$
(\rho_{\cal A} \N \rho_{\cal A} ) = (\rho_{\cal A} \p \rho_{\cal A} ) + \Pi^A \Jb^I A_{IA} ,$$
where $\Jb^I=\ha K^I_{\cal A\cal B} \rho_{\cal A}\rho_{\cal B}$ with $K^I$ represents the generators of the Lie algebra of $E_8 \times E_8$ or $SO(32)$ in the fundamental representation, and $A_{IA}$ is the corresponding gauge field. Finally, $S_{FT}$ is the Fradkin-Tseytlin term given by the world-sheet integral of the dilaton superfield $\Phi$.

In  \BerkovitsUE\ it was shown that the charge $Q_0$ in \quis\ is nilpotent and conserved if the background is constrained to satisfy the supergravity and SYM equations of motion in ten dimensions.  Alternatively, in \ChandiaXI\ it was found how the world-sheet fields of the action \scurvmin\ transform under $Q_0$ and it was verified that the action $S_0$ is BRST invariant. We will assume that the minimal variables are unaffected by the non-minimal BRST charge. Below, we will need the BRST transformation of the connection $\Ob_\a{}^\b$ which is equal to
\eqn\qom{ Q_0 \Ob_\a{}^\b = \Nb ( \l^\g \O_{\g\a}{}^\b ) - \l^\g \Pib^A R_{A\g\a}{}^\b ,}   
where the covariant derivative was defined in \covder\ and $R$ is the curvature superfield. Note that the first term in this transformation is a Lorentz transformation with the field-dependent parameter $\l^\g \O_{\g\a}{}^\b$. In \ChandiaXI\ it was shown that this property is true for all the world-sheet fields of the action \scurvmin, that is, the BRST transformation of the fields always contains a gauge and Lorentz transformation. We will denote as $\Qt_0$ on the world-sheet fields as the minimal pure spinor BRST transformation without including the Lorentz rotation. Below we will need the action of $\Qt_0$ on the other world-sheet fields. These transformations were derived in \ChandiaXI\  and the non-vanishing variations are  
\eqn\qtonws{ \Qt_0 \Pi^A = \d_\a^A \N \l^\a - \l^\a \Pi^B T_{B\a}{}^A,\quad \Qt_0 \o_\a = d_\a , }
$$
\Qt_0 d_\a = \l^\b \Pi^a (\g_a)_{\b\a} + \l^\b \l^\g \o_\d R_{\a\b\g}{}^\d ,$$
where we have used some local Lorentz symmetry to gauge fix the torsion component $T_{\a\b}{}^\g$ to zero \BerkovitsUE.

\newsec{ BRST Transformations of the Non-minimal Variables}

One could think that the $Q_0$ BRST transformations of the non-minimal variables are trivial. However, all the fields in the minimal model transform, at least, with a Lorentz rotation term. Since the non-minimal variables transform under Lorentz rotations, then we expect a non trivial action of $Q_0$ on the non-minimal pure spinor variables.  One could argue that the Lorentz index in the non-minimal variables just counts number of fields and it is not a vector index. This is not the case. The $b$ ghost in flat space is constructed from contractions between minimal and non-minimal variables such that it is Lorentz invariant. Using cohomology arguments, we will find the form in which $Q_0$ acts on the non-minimal variables.

The BRST charge in a curved background has the same form as in flat space,
that is,
\eqn\brstc{ Q = Q_0 + Q_1 ,}
where both $Q_0$ and $Q_1$ are nilpotent. Because $Q$ is nilpotent, we obtain the $Q_0$ and $Q_1$ anticommute.  The action of $Q_1$ on the non-minimal variables is
\eqn\qonenm{ Q_1 \lh_\a = - r_\a,\quad Q_1 s^\a = \oh^\a,\quad Q_1 \oh^\a = Q_1 r_\a = 0 .}
Consider the last two equations here. Acting with $Q_0$ and using anticommutation with $Q_1$, we obtain
\eqn\qzeronma{ÊQ_1 ( Q_0 \oh^\a ) = 0,\quad Q_1 ( Q_0 r_\a ) = 0 .}
Because the cohomology of $Q_1$ is trivial \BerkovitsBT, we determine the form of $Q_0 \oh^\a$ and $Q_0 r_\a$ to be
\eqn\getq{ Q_0 \oh^\a = \oh^\b {\cal A}_\b{}^\a +r_\b \GG^{\b\a},\quad Q_0 r_{\a} = \BB_\a{}^\b r_\b + \oh^\b \HH_{\b\a} ,}
where $\AA, \BB, \GG, \HH$ depend on the minimal variables. Note that the terms involving $\AA$ and $\BB$ contains Lorentz rotations, so we identify them with the field-dependent Lorentz parameter of the minimal sector. Recall that $Q_0$ maps a $\l$-ghost number $n$ field to a $\l$-ghost number $n+1$ field, so we need to add a factor of $\l$ to the above transformations. Then, we have
\eqn\qzeroor{ÊQ_0 \oh^\a = -\oh^\b  \l^\g ( \O_{\g\b}{}^\a + \AA_{\g\b}{}^\a ) +\l^\b r_\g \GG_\b{}^{\g\a},\quad Q_0 r_{\a} = \l^\g ( \O_{\g\a}{}^\b + \BB_{\g\a}{}^\b ) r_\b + \l^\b\oh^\g \HH_{\b\g\a} ,}
where $\AA, \BB, \GG$ and $\HH$ depend on the minimal variables only. Note that $\HH$ has conformal weight $-1$, then it has to vanish because it is not possible to write a quantity of such conformal weight in the minimal formalism, at least if one requires Lorentz covariance. Let us try the possibility that both $\GG$ and $\HH$ vanish. Below, we will determine $\AA$ and $\BB$ by requiring that $Q_0$ is nilpotent.  Before that, we obtain the action of $Q_0$ on the other non-minimal variable $\lh$ and $s$. Analogous to \qzeronma\ we have
\eqn\qzeronmb{ Q_1 ( Q_0 \lh_\a ) = Q_0 r_\a,\quad Q_1 ( Q_0 s^\a ) = - Q_0 \o^\a ,}
from which we obtain
\eqn\qzerols{ Q_0 \lh_\a = \l^\g ( \O_{\g\a}{}^\b + \BB_{\g\a}{}^\b ) \lh_\b,\quad Q_0 s^\a = s^\b \l^\g ( \O_{\g\b}{}^\a + \AA_{\g\b}{}^\a ) .}  

Because $Q_0$ is nilpotent, $\AA$ and $\BB$ in \qzeroor\ and \qzerols\ satisfy certain constraints. Applying $Q_0$ to first equation in \qzeroor, we obtain
\eqn\qtwoom{ Q_0^2 \oh^\a = -\ha \oh^\b \l^\g \l^\d ( R_{\g\d\b}{}^\a + \N_{(\g} \AA_{\d)\b}{}^\a - \AA_{\g\b}{}^\r \AA_{\d\r}{}^\a - \AA_{\d\b}{}^\r \AA_{\g\r}{}^\a ) ,}
where we used the definition of the curvature $R$ in terms of the connection $\O$ and the Berkovits-Howe constraint $\l^\g \l^\d T_{\g\d}{}^A = 0$, where $T$ is the torsion in superspace \BerkovitsUE. The solution for $\AA$ is
\eqn\valueA{ÊA_{\g\b}{}^\a = -{1\over 4} (\g^{ab})_\b{}^\a T_{\g ab} .}
To verify that \valueA\ is the solution of \qtwoom, we use 
$$
R_{\g\d\b}{}^\a = R_{\g\d} \d_\b^\a + {1\over 4} (\g^{ab})_\b{}^\a R_{\g\d ab} ,$$
where $R_{\g\d} = \N_{(\g} \O_{\d)}$. Note that $\l^\g \l^\d R_{\g\d}$ vanishes after using that $\O_\a$ is proportional to $\N_\a \Phi$, where $\Phi$ is the dilaton superfield \BerkovitsUE.  The last step is to use the Bianchi identity involving $R_{\g\d ab}$ and, again, the constraint $\l^\g \l^\d T_{\g\d}{}^A = 0$.

We proceed  similarly to determine $\BB$ by demanding that $Q_0^2 r_\a$ vanishes. It turns out that $\BB = \AA$. In summary, the action of $Q_0$ on the non-minimal variables is given by
\eqn\qzeroonNM{ \eqalign{& Q_0 \oh^\a = - \oh^\b \l^\g \left( \O_{\g\b}{}^\a - {1\over 4} (\g^{ab})_\b{}^\a T_{\g ab} \right),\cr & Q_0 r_\a = - \l^\g r_\b \left( \O_{\g\a}{}^\b - {1\over 4} (\g^{ab})_\a{}^\b T_{\g ab} \right), \cr &
Q_0 \lh_\a = \l^\g \lh_\b \left( \O_{\g\a}{}^\b - {1\over 4} (\g^{ab})_\a{}^\b T_{\g ab} \right), \cr & Q_0 s^\a = s^\b \l^\g \left( \O_{\g\b}{}^\a - {1\over 4} (\g^{ab})_\b{}^\a T_{\g ab} \right). \cr}}

\newsec{ The Non-minimal Heterotic Pure Spinor String on a Curved Background}

We now define a world-sheet action for the heterotic pure spinor string in a curved background. We start with the generalization
of \Qequiv\ in this case,
\eqn\Scurved{ÊS = S_0 + Q \int d^2z ~ s^\a \Nb \lh_\a ,}
where $S_0$ is given in \scurvmin. Using the above BRST transformations on the minimal and non-minimal variables we obtain
\eqn\SNM{  \eqalign { S &= S_0 + \int d^2z ~ \oh^\a \Nb \lh_\a + s^\a \Nb r_\a \cr &+ {1\over 4} s^\b \lh_\g \Nb \l^\a (\g^{ab})_\b{}^\g T_{\a ab} + \l^\a  s^\b \lh_\g \Pib^A ( {1\over 4} (\g^{ab})_\b{}^\g \N_A T_{\a ab} - R_{A\a\b}{}^\g ) .\cr } }
Because the pure spinor BRST operator is nilpotent on gauge invariant operator, the action \Scurved\ is BRST invariant.

Since we are interested in the construction of the $b$ ghost, we need to know the left-moving stress-energy tensor derived from \SNM. Under an holomorphic conformal transformation, the superspace coordinate $Z$, the pure spinor ghosts $\l, \lh$ and $r$ carry conformal weight zero. While the conjugate pure spinor variables $\o, \oh$ and $s$ carry conformal weight one. Noether theorem determines the conserved charge due to this transformation to be
\eqn\TT{ \eqalign{T &= T_0 - \oh^\a \N \lh_\a - s^\a \N r_\a \cr & - {1\over 4}Ês^\b \lh_\g \N \l^\a (\g^{ab})_\b{}^\g T_{\a ab} - \l^\a s^\b \lh_\g  \Pi^A ( {1\over 4} (\g^{ab})_\b{}^\g \N_A T_{\a ab} - R_{A\a\b}{}^\g ) ,\cr }}
where $T_0$ is the stress-tensor from the action $S_0$ which is given by
\eqn\Tmin{ T_0 = - \ha \Pi_a \Pi^a - d_\a \Pi^\a - \o_\a \N \l^\a .}
Note that \TT\ reduces to the correct expression in the flat space limit because 
$$
T_{\a ab} = 2 (\g_{ab})_\a{}^\b \O_\b \to 0, \quad R_{A\a\b}{}^\g \to 0 .$$

In the next section we will use the stress tensor \TT\ to determine the $b$ ghosts satisfying $Q b = T$.

\newsec{ The $b$ ghost}

The $b$ ghost was constructed in \BerkovitsBT\ for the case in which the background is flat. It is given by
\eqn\bflat{ b = - s^\a \p \lh_\a + {1\over(\l\lh)} \lh_\a G^\a - {1\over(\l\lh)^2} \lh_\a r_\b H^{\a\b} - {1\over(\l\lh)^3} \lh_\a r_\b r_\g K^{\a\b\g} + {1\over(\l\lh)^4} \lh_\a r_\b r_\g r_\d L^{\a\b\g\d} ,}
where $G, H, K$ and $L$ have conformal weight two, $\l$-ghost number zero and depend on the minimal variables.  They have to satisfy the relations
\eqn\GHKL{ÊQ G^\a = \l^\a T,\quad Q H^{\a\b} = \l^{[\a} G^{\b]},\quad Q K^{\a\b\g} = \l^{[\a} H^{\b\g]},\quad Q L^{\a\b\g\d} = \l^{[\a} K^{\b\g\d]},\quad \l^{[\a} L^{\b\g\d\r]} = 0 .}
to satisfy $Qb=T$. The expressions for them were derived in \BerkovitsBT, we quote the result here
\eqn\GHKLflat{ G^\a = -\ha \Pi^m \g_m^{\a\b} d_\b - {1\over4} J \p\t^\a + {1\over8} N^{mn} (\g_{mn}\p \t)^\a ,}
$$
H^{\a\b} = {1\over192} \left( (d\g^{mnp} d) + 24 \Pi^m N^{np} \right) \g_{mnp}^{\a\b} ,$$
$$
K^{\a\b\g} = {1\over16} \g_{mnp}^{[\a\b}(\g^m d)^{\g]} N^{np} ,\quad L^{\a\b\g\d} = {1\over128} \g_{mnp}^{[\a\b} (\g^{pqr})^{\g\d]} N^{mn} N_{qr} ,$$ 
where  $\Pi^m = \p X^m + \ha (\t\g^m\p\t)$ is the supersymmetric world-sheet momentum, $J=-\l^\a \o_\a$ is the $\l$-ghost number current, and $N^{mn}=\ha(\l\g^{mn}\o)$ is the generator for Lorentz transformations of the pure spinor variables.

We generalize the expression \bflat\ to a curved background and find the relations analogous to \GHKL. We propose
\eqn\bcurv{Êb = - s^\a \N \lh_\a + {1\over(\l\lh)} \lh_\a G^\a - {1\over(\l\lh)^2} \lh_\a r_\b H^{\a\b} - {1\over(\l\lh)^3} \lh_\a r_\b r_\g K^{\a\b\g} + {1\over(\l\lh)^4} \lh_\a r_\b r_\g r_\d L^{\a\b\g\d} ,}
and now we compute $Qb$ and impose it is equal to \TT. Because of the form in which $Q$ acts on both minimal and no-minimal variables, we can organize $Qb$ in an expansion in powers of $r$. Note that the $b$ ghost is a Lorentz scalar, so all the terms which depend on the Lorentz connection $\O$ will produce a zero variation of $b$ because they are a field-dependent Lorentz transformation.  

The term independent from $r$ is 
\eqn\rzero{ T - T_0  + {1\over(\l\lh)} \lh_\a \left( \Qt_0 G^\a - {1\over4} \l^\b T_{\b ab} (\g^{ab})_\g{}^\a G^\g - 5 \l^\b \O_\b G^\a \right)  ,}
where $\Qt_0 G^\a$ is the minimal BRST transformation without the Lorentz rotation term. It is required that
\eqn\eqG{ \Qt _0 G^\a = \l^\a T_0 +  {1\over4} \l^\b T_{\b ab} (\g^{ab})_\g{}^\a G^\g + 5 \l^\b \O_\b G^\a .}
In this way, \rzero\ is equal to the stress-energy tensor $T$. The remaining terms in $Qb$ has to vanish and we find constraints equations for $H, K, L$. The order $1$ in $r$ determines an equation for $H$,
\eqn\eqH{ \Qt_0 H^{\a\b} = \l^{[\a} G^{\b]} - {1\over4} \l^\g T_{\g ab} (\g^{ab})_\d{}^{[\a} H^{\b]\d} + 10 \l^\g \O_\g H^{\a\b} .}
The order $2$ in $r$ determines an equation for $K$,
\eqn\eqK{ \Qt_0 K^{\a\b\g} = \l^{[\a} H^{\b\g]} + {1\over4} \l^\d T_{\d ab} (\g^{ab})_\r{}^{[\a} K^{\b\g]\r} + 15 \l^\d \O_\d K^{\a\b\g} .}
The order $3$ in $r$ determines an equation for $L$,
\eqn\eqL{ \Qt_0 L^{\a\b\g\d} = \l^{[\a} K^{\b\g\d]} - {1\over4} \l^\r T_{\r ab} (\g^{ab})_\s{}^{[\a} L^{\b\g\d]\s} + 20 \l^\r \O_\r L^{\a\b\g\d} .}
Finally, the order $4$ in $r$ determines the constraint for $L$,
\eqn\constL{ \l^{[\a} L^{\b\g\d\r]} = 0 .}
Note that, after adding the Lorentz rotation term, the nilpotency of $Q_0$ on $G, H, K, L$ is verified. Note that the equations \eqG\ to \constL\ have the correct flat space limit because $\O_\a=0$ in this case. 

\newsec{ Construction of the $b$ Ghost}

We look for the fields $(G, H, K, L)$ satisfying equations \eqG\ to \constL. They have conformal dimension two and minimal ghost number zero. Quite general, they all have the form
\eqn\GHKL{ U^A = \Pi^a \Pi^b (u_1)_{ab}{}^A + \Pi^a \Pi^\b (u_2)_{a\b}{}^A + \Pi^a d_\b (u_3)_a{}^{\b A} + \Pi^a \l^\b \o_\g (u_4)_{a\b}{}^{\g A} + \Pi^\b \Pi^\g (u_5)_{\b\g}{}^A }
$$ + \Pi^\b d_\g (u_6)_\b{}^{\g A} + \Pi^\b \l^\g \o_\d (u_7)_{\b\g}{}^{\d A} + d_\b d_\g (u_8)^{\b\g A} + d_\b \l^\g \o_\d (u_9)_\g{}^{\d\b A} + \l^\b \o_\g \l^\d \o_\r (u_{10})_{\b\d}{}^{\g\r A} ,$$
where the $u$'s are super fields of the background and the index $^A$ is $^\a$ for $G$, $^{\a\b}$ for $H$, $^{\a\b\g}$ for $K$ and $^{\a\b\g\d}$ for $L$. There are possible terms involving $\N\Pi^A, \N d_\a, \N\l^\a, \N\o_\a$. We will not need these terms, so we do not include them. Note that all the terms involving $\o$ must be invariant under the gauge transformation $\d \o_\a = (\l\g^a)_\a \L_a$. It constrains the Lorentz index structure of these terms above. 
 
We need to know the action of $\Qt_0$ on $G, H, K, L$ to solve the equations \eqG\ to \constL. Then, we compute $\Qt_0$ on the general world-sheet field $U^A$. Using \qtonws\ we obtain
\eqn\QtU{Ê\Qt_0 U^A =  \l^\b \Pi^a \Pi^b \left( -2 T_{a\b}{}^c (u_1)_{cb}{}^A + \N_\b (u_1)_{ab}{}^A + (\g_a)_{\b\g} (u_3)_b{}^{\g A} \right) }
$$
+ \l^\b \Pi^a \Pi^\g \left( -2 T_{\g\b}{}^b(u_1)_{ba}{}^a + T_{a\b}{}^b (u_2)_{b\g}{}^A - \N_\b (u_2)_{a\g}{}^A - (\g_a)_{\b\d} (u_6)_\g{}^{\d A} \right)$$ $$
+ \l^\b \Pi^\g \Pi^\d \left( - T_{\g\b}{}^a (u_2)_{a\d}{}^A + \N_\b (u_5)_{\g\d}{}^A \right) 
+ \N\l^\b \Pi^a  (u_2)_{a\b}{}^A $$
$$
+ \l^\b d_\g \Pi^a \left( T_{a\b}{}^b (u_3)_b{}^{\g A} - \N_\b (u_3)_a{}^{\g A} + (u_4)_{a\b}{}^{\g A} + 2 (\g_a)_{\b\d} (u_8)^{\d\g A} \right)$$
$$
+ \l^\b d_\g \Pi^\d \left( T_{\d\b}{}^a (u_3)_a{}^{\g A} - \N_\b (u_6)_\d{}^{\g A} + (u_7)_{\d\b}{}^{\g A} \right)$$
$$
+ \l^\b \l^\g \o_\d \Pi^a \left( R_{\r\b\g}{}^\d (u_3)_a{}^{\r A} - T_{a\b}{}^b (u_4)_{b\g}{}^{\d A} + \N_\b (u_4)_{a\g}{}^{\d A} + (\g_a)_{\b\r} (u_9)_\g{}^{\d\r A} \right) $$ 
$$
+ \l^\b \l^\g \o_\d \Pi^\r \left(- T_{\r\b}{}^a (u_4)_{a\g}{}^{\d A} - R_{\s\b\g}{}^\d (u_6)_\r{}^{\s A} - \N_\b (u_7)_{\r\g}{}^{\d A} \right) $$ $$
+  2 \N\l^\b \Pi^\g  (u_5)_{\b\g}{}^A 
+ \N\l^\b d_\g  (u_6)_\b{}^{\g A}  + \N\l^\b \l^\g \o_\d  (u_7)_{\b\g}{}^{\d A}  
+ \l^\b d_\g d_\d \left( \N_\b (u_8)^{\g\d A} - (u_9)_\b{}^{\d\g A} \right) $$
$$
+ \l^\b \l^\g \o_\d d_\r \left( 2 R_{\s\b\g}{}^\d (u_8)^{\s\r A} - \N_\b (u_9)_\g{}^{\d\r A} + 2 (u_{10})_{\b\g}{}^{\r\d A} \right) $$
$$
+ \l^\b \l^\g \o_\d \l^\r \o_\s \left( \N_\b (u_{10})_{\g\r}{}^{\d\s A} + R_{\tau\b\g}{}^\d (u_9)_\r{}^{\s\tau A} \right),$$
where $T$ is the torsion, $R$ is the curvature, both in superspace. Now we solve the equations to obtain the $b$ ghost. We well call the $u$'s superfields in \GHKL\ $g$ for $G$, $h$ for $H$, $k$ for $K$ and $l$ for $L$. 

We now solve the equation for $G^\a$ \eqG. The expression for $\Qt_0 G^\a$ can be read from \QtU\ and it has to equal to the right hand side of \eqG.  The rhs of \eqG\ does not contain a term  involving $\N\l^\b \Pi^\g$, then,  $(g_5)_{\b\g}{}^\a=0$. The term with $\N \l^\b \Pi^a$ in \eqG\ implies that $(g_2)_{a\b}{}^\a = 0$. The term with $\N\l^\a d_\g$ in \eqG\ implies that $(g_6)_\b{}^{\g\a} = 0$.

Consider the term involving $\l^\b \Pi^a \Pi^\g$ in the equation \eqG. It determines
\eqn\eqgh{Ê\g^b_{\g\b} (g_1)_{ba}{}^\a = 0 .}
If we multiply this equation by $\g_c^{\b\g}$, we obtain that $(g_1)_{ca}{}^\a = 0$. 

Consider now the term involving $\l^\b \Pi^a \Pi^b$ in the equation \eqG. Because $g_1$ is zero, the rhs vanishes and we have
\eqn\eqgthree{ (\g_{(	a})_{\b\g} (g_3)_{b)}{}^{\g\a} = -Ê\eta_{ab} \d_\b^\a ,}
which implies
\eqn\solgthree{ (g_3)_{a}{}^{\b\a} = - \ha \g_a^{\b\a} .}

Consider the term involving $\l^\b d_\g \Pi^a$ in \eqG. It leads to the equation
\eqn\gfge{Ê(g_4)_{a\b}{}^{\g\a} + 2(\g_a)_{\b\d} (g_8)^{\d\g\a} = {3\over2} \g_a^{\g\a} - \ha (\g_{ab})_\b{}^\r (\g^b)^{\g\a} \O_\r + {1\over4} \g_{abc}^{\g\a} (\g^{bc})_\b{}^\r \O_\r .}
To solve this equation we note that $g_4$ contains a $0$-form and a $2$-form when it is expanded in $_\b{}^\g$, and  $g_8$ is antisymmetric in $^{\d\g}$. That is,
\eqn\gsexp{ (g_4)_{a\b}{}^{\g\a} = \d_\b^\g (j_a^\a) + (\g^{bc})_\b{}^\g (j_a^\a)_{bc},\quad (g_8)^{\d\g\a} = \g_{abc}^{\d\g} (k^\a)^{abc} .}
Plugging \gsexp\ into \gfge\ we obtain
\eqn\solvn{ \d_\b^\g (j_a^\a) + (\g^{bc})_\b{}^\g ( (j_a^\a)_{bc} + 6 (k^\a)_{abc} ) + 2 (\g_{abcd})_\b{}^\g (k^\a)^{bcd} = {3\over2} \g_a^{\g\a} - \ha (\g_{ab})_\b{}^\r (\g^b)^{\g\a} \O_\r + {1\over4} \g_{abc}^{\g\a} (\g^{bc})_\b{}^\r \O_\r .}
Multiplying by $\d_\g^\b$ we determine the $0$-form of $g_4$, 
\eqn\jzero{ (j_a^\a) = {3\over2}\g_a^{\a\b} \O_\b .}
Multiplying by $(\g^{aefg})_\g{}^\b$ we obtain that $g_8$ vanishes. Finally, multiplying by $(\g^{de})_\g{}^\b$ we obtain 
\eqn\jtwo{ (j^\a_a)_{bc} = - {1\over4} (\g_a\g_{bc})^{\a\b} \O_\b .}
In summary, the solution of \gfge\ is
\eqn\solgfge{ (g_4)_{a\b}{}^{\g\a} = {3\over2} \d_\b^\g \g_a^{\a\d} \O_\d - {1\over4} (\g^{bc})_\b{}^\g (\g_a \g_{bc})^{\a\d} \O_\d ,\quad (g_8)^{\d\g\a} = 0 .}

Because $g_8$ vanishes, it seems that  $g_9$ and $g_{10}$ also vanish in order to simplify the equations from \eqG. In this case, it remains to determine $g_7$. It is obtained from the equation involving $\l^\b d\g \Pi^\d$. In fact
\eqn\gseven{Ê(g_7)_{\d\b}{}^{\g\a} = -\d_\b^\a \d_\d^\g + \ha \g^a_{\d\b} \g_a^{\g\a} .} 
Note that $g_7$ satisfies 
\eqn\gsevright{Ê(\g^{abcd})_\g{}^\b (g_7)_{\d\b}{}^{\g\a} = 0 ,}
as it is required from the gauge symmetry for $\o$ in the minimal pure spinor string. Note that the term involving $\N\l^\b \l^\g \o_\d$ in \eqG\ is satisfied because $(\N\l \g^a \l) = 0$.

It remains to verify that the terms involving $\l^\b \l^\g \o_\d \Pi^a$ and $\l^\b \l^\g \o_\d \Pi^\r$. Instead of plugging the values of the $g$ superfields that we have determined. We will show that these equations are implied by the others. Consider the first term. It implies the equation
\eqn\llpia{ \l^\b \l^\g \left( \N_\b (g_4)_{a\b}{}^{\d\a} + T_{\b a}{}^b (g_4)_{b\g}{}^{\d\a} + R_{\r\b\g}{}^\d (g_3)_a{}^{\r\a} \right) }
$$
= \l^\b \l^\g \left( {1\over4} T_{\b bc} (\g^{bc})_\r{}^\a (g_4)_{a\g}{}^{\d\r} + 5 \O_\b (g_4)_{a\g}{}^{\d\a} \right) .$$
We will show that this equation is implied by the term involving $\l^\b d_\g \Pi^a$ which states that 
\eqn\ldpia{Ê(g_4)_{a\g}{}^{\d\a} = \N_\g (g_3)_a{}^{\d\a} + T_{\g a}{}^b (g_3)_b{}^{\d\a} - {1\over4} T_{\g bc} (\g^{bc})_\r{}^\a (g_3)_a{}^{\d\r} - 5 \O_\g (g_3)_a{}^{\d\a} .}
We act with $\N_\b$ on this equation and symmetrize in $(\b\g)$ to obtain
\eqn\llpiaa{Ê\N_{(\b} (g_4)_{a\g)}{}^{\d\a} + T_{(\b a}{}^b (g_4)_{b\g)}{}^{\d\a} - {1\over4} T_{(\b bc} (\g^{bc})_\r{}^\a (g_4)_{a\g)}{}^{\d\r} - 5 \O_{(\b} (g_4)_{a\g)}{}^{\d\a} }
$$
= \{ \N_\b , \N_\g \} (g_3)_a{}^{\d\a} + \left( \N_{(\b} T_{\g) a}{}^b - T_{a(\b}{}^c T_{\g)c}{}^b \right) (g_3)_b{}^{\d\a} - 5 \N_{(\b} \O_{\g)} (g_3)_a{}^{\d\a} $$ $$- {1\over4}Ê \left(  (\g^{bc})_\r{}^\a \N_{(\b} T_{\g) bc} - {1\over4} (\g^{de} \g^{bc})_\r{}^\a T_{(\b bc} T_{\g)de} \right) (g_3)_a{}^{\d\r} .$$ 
where the symmetrization is on $(\b\g)$ only. Note that we will multiply this expression by $\l^\b\l^\g$ to obtain \llpia. Then, the last term in the second  line will vanish because $\O_\a$ is proportional to $\N_\a \Phi$, where $\Phi$ is the dilation superfield. Recall that the anticommutator in the second line is related to the curvature. In fact, the graded commutator for covariant derivatives on $g_3$ is
\eqn\antic{ [ \N_A , \N_B ] (g_3)_a{}^{\d\a} = - T_{BA}{}^C \N_C (g_3)_a{}^{\d\a} + (g_3)_a{}^{\s\a} R_{BA\s}{}^\d + (g_3)_a{}^{\d\s} R_{BA\s}{}^\a - R_{BAa}{}^b (g_3)_b{}^{\d\a} .}
Using this equation in \llpiaa\ and the Bianchi identity involving the curvature $R_{\b\g a}{}^b$ we obtain
\eqn\llpiab{ \N_{(\b} (g_4)_{a\g)}{}^{\d\a} + T_{(\b a}{}^b (g_4)_{b\g)}{}^{\d\a} - {1\over4} T_{(\b bc} (\g^{bc})_\r{}^\a (g_4)_{a\g)}{}^{\d\r} - 5 \O_{(\b} (g_4)_{a\g)}{}^{\d\a} - R_{\b\g\r}{}^\d (g_3)_a{}^{\r\a} }
$$
=  - {1\over4}Ê \left( (\g^{bc})_\r{}^\a \N_{(\b} T_{\g) bc} - {1\over4} (\g^{de} \g^{bc})_\r{}^\a T_{(\b bc} T_{\g)de} - (\g^{bc})_\r{}^\a R_{\b\g bc} \right) (g_3)_a{}^{\d\r}$$
up to terms proportional to $\g^b_{\b\g}$ which will be zero after hitting with $\l^\b\l^\g$. In the lhs here is equal to the lhs of \llpia, after multiplying with $\l^\b\l^\g$, if we use the Bianchi identity $R_{(\b\g\r)}{}^\d = 0$. The rhs here is because the Bianchi identity for $R_{\b\g bc}$. Therefore, we have proved that equation \llpia\ is implied by the one of the other equations. A similar calculation determines that the term involving $\l^\b \l^\g \o_\d \Pi^\r$ in \eqG\ is satisfied. 

In summary, we determine $G^\a$ to be
\eqn\defG{ G^\a = - \ha \Pi^a d_\b \g_a^{\b\a} - {1\over4} J \Pi^\a + {1\over8} N^{ab} (\g_{ab})^\a{}_\b \Pi^\b - {3\over2} J \Pi^a \g_a^{\a\b} \O_\b - \ha N^{ab}Ê\Pi^c (\g_c \g_{ab} )^{\a\b} \O_\b ,}
where $J=-\l^\a\o_\a$ and $N^{ab} = \ha (\l \g^{ab} \o)$. Note that this expression has the correct flat space limit. In this case
\eqn\limG{Ê\Pi^\a \to \p\t^\a,\quad \O_\a \to 0 .} 

Consider now the equation for $H^{\a\b}$ \eqH.  Note that the rhs here does not contain terms with derivatives of $\l^\g$, therefore $h_2 = h_5 = h_6 = h_7 = 0$.  Consider the term with $\l^\g \Pi^a \Pi^\d$, it implies that $h_1=0$. Consider now the term with $\l^\g \Pi^a \Pi^b$ that implies \eqn\htres{ (\g_{(a})_{\g\d} (h_3)_{b)}{}^{\d\a\b} = 0 .} 
Multiplying by $\eta^{ab}$ we obtain
$$
\g^a_{\g\d} (h_3)_a{}^{\d\a\b} = 0 .$$
And multiplying \htres\ by $(\g^b)^{\s\g}$ and we use the above restriction on $h_3$ we obtain the it vanishes.  

The term with $\l^\g d_\d \Pi^a$ determines $h_4$ and $h_8$. In fact, the corresponding term gives the equation
\eqn\eqhfhe{Ê(h_4)_{a\g}{}^{\d\a\b} + 2 (\g_a)_{\g\r} (h_8)^{\r\d\a\b} = -\ha \d_\g^{[\a} \g_a^{\b]\d} .}
 Note that the four form in the expansion of $h_4$ in $_\g{}^\d$ vanishes and that $h_8$ is antisymmetric in $^{\r\d}$. Then,
 $$(h_4)_{a\g}{}^{\d\a\b} = \d_\g^\d (x_a^{\a\b}) + (\g^{bc})_\g{}^\d (x_a^{\a\b})_{bc},\quad (h_8)^{\r\d\a\b} = \g_{bcd}^{\r\d} (y^{\a\b})^{bcd} .$$
 Plugging these expressions into \eqhfhe\ and multiplying by $\d_\d^\g$, then by $(\g^{aefg})_\d{}^\d$, and finally by $(\g^{de})_\d{}^\g$ we determine $h_4$ and $h_8$. They are
 \eqn\hfhe{ (h_4)_{a\g}{}^{\d\a\b} = -{1\over{16}} (\g^{bc})_\g{}^\d \g_{abc}^{\a\b},\quad (h_8)^{\r\d\a\b} = {1\over{192}} \g_{abc}^{\r\d} (\g^{abc})^{\a\b} .}

Consider the equation determined by the term with $\l^\g d_\d d_\r$ in \eqH. It determines the part of $h_9$ antisymmetric in $^{\r\d}$ to be
\eqn\ashnine{ (h_9)_\g{}^{[\r\d]\a\b} = {1\over{16}} (\g^{abc})^{\a\b} (\g_{dbc})^{\r\d} (\g^d\g_a)_\g{}^\s \O_\s .}
Note that the remaining equations in \eqH\ are satisfied if $h_{10}=0$. As in the case for $G^\a$, the equations from \eqH\ with curvature are implied by the equations without curvature and the use of Bianchi identities and the pure spinor condition. It remains to determine the symmetric part in $^{\r\d}$ of $h_9$. It turns out that it can be expanded as
\eqn\hnsymm{ (h_9)_\g{}^{(\r\d)\a\b} = \left( H^1_{\g abcd} (\g^d)^{\r\d} + H^5_{\g abcdefgh} (\g^{defgh})^{\r\d} \right) (\g^{abc})^{\a\b} .}
The condition $(\g^{ijkl})_\r{}^\d (h_9)_\g{}^{\r\d\a\b} = 0$ determines the equation, 
\eqn\Hone{Ê(\g^d \g_{ijkl})^{\d\g} H^1_{\g abcd} + (\g^{defgh} \g_{ijkl})^{\d\g} H^5_{\g abcdefgh} + {1\over16} (\g_{dbc} \g_{ijkl} \g^d \g_a)^{\d\g} \O_\g = 0 .}
Note that this equation has to be completely antisymmetric in $^{abc}$ because we factor out the matrix $(\g^{abc})^{\a\b}$. Because of this, we try the solution
\eqn\solH{ H^1_{\g abcd} = \left( A (\g_{abc} \g_d)_\g{}^\s + B (\g_d \g_{abc})_\g{}^\s \right) \O_\s, }
$$
H^5_{\g abcd} = \left( C (\g_{abc} \g_{defgh})_\g{}^\s + D (\g_{defgh} \g_{abc})_\g{}^\s \right) \O_\s.$$
The constants $A,B,C,D$ can be determined when we plug this solution into \Hone. 

Up to these constants, $H^{\a\b}$ is given by
\eqn\defH{ H^{\a\b} = {1\over8}Ê \Pi^a N^{bc} \g_{abc}^{\a\b} + {1\over192} (d\g_{abc}d) \g_{abc}^{\a\b} + d_\g \l^\d \o_\r (h_9)_\d{}^{\r\g\a\b} ,}
where $h_9$ is given above, up to some undetermined constants, and depends on $\O_\a$. Therefore, the flat limit of \defH\ gives the expected result because $h_9 \to 0$.

We proceed similarly to determine $K$ and $L$. The calculation becomes more involving. We just can state that $K$ has the form
\eqn\solK{ K^{\a\b\g} = K_0^{\a\b\g} + \l^\d \o_\r \l^\s \o_\tau (k_{10})_{\d\s}{}^{\r\tau\a\b\g} ,}
where $K_0$ is the value of $K$ in the flat space limit and $k_{10}$ depends linearly on $\O_\a$ so it becomes zero in the flat space limit. Finally, the tensor $L$ is equal to the corresponding tensor in flat space-time background. 

\appendix{A}{ Review on Pure Spinor Superspace}

We review the results from \BerkovitsUE. The string action \scurvmin\ is based on the superspace coordinate $Z^M$, where $M$ is a target space super index and runs over ten bosonic indices and sixteen fermionic indices. We define the world-sheet fields $\Pi^A$ and $\Pib^A$ as in \pis\ by introducing the supervielbein $E_M{}^A$, where $A$ is a local superspace index. We also need a super connection  $\O_{M A}{}^B$ to write super covariant derivatives. Out of $E_M{}^A$ and $\O_{M A}{}^B$ we define the super one-forms
\eqn\EO{ÊE^A = dZ^M E_M{}^A,\quad \O_A{}^B = dZ^M \O_{M A}{}^B .} 
We can define now a covariant derivative in superspace which transform homogeneously under local Lorentz rotation. For a super $p$-form $\Psi^A$ it is given by
\eqn\covD{ \N \Psi_B{}^A =  d \Psi_B{}^A + \Psi_B{}^C \O_C{}^A - (-1)^p \O_B{}^C \Psi_C{}^A .}  
In this formula the product between forms is a wedge product. Given the forms \EO\ and the derivative \covD\ we define the super two forms torsion $T^A$ and curvature $R_B{}^A$ as
\eqn\TR{ÊT^A = \N E^A,\quad R_B{}^A = d\O_B{}^A + \O_B{}^C \O_C{}^A .}
They satisfy the Bianchi identities
\eqn\bianchi{Ê\N T^A = T^B R_B{}^A,\quad \N R_B{}^A = 0 .}
We use these identities in the torsion and curvature components,
\eqn\comp{ T^A = \ha E^B E^C T_{CB}{}^A,\quad R_B{}^A = \ha E^C E^D R_{DCB}{}^A .}
In terms of the torsion and curvature components, the Bianchi identities \bianchi\ become
\eqn\bianchiC{ \N_{[A} T_{BC]}{}^D + T_{[AB}{}^E T_{EC]}{}^D - R_{[ABC]}{}^D = 0,\quad \N_{[A} R_{BC]D}{}^E + T_{[AB}{}^F R_{FC]D}{}^E = 0.}

In \BerkovitsUE\ and \ChandiaXI, the BRST invariance of the action \scurvmin\ puts the background on-shell. In fact, the nil potency of $Q$ implies $\l^\a \l^\b T_{\a\b}{}^A = 0$. Berkovits and Howe showed that Lorentz invariance and a symmetry involving the pure spinor variables and the connection $\O$ ( that they call shift symmetry) allow to fix the values of the torsion component as
\eqn\torscomp{ÊT_{\a\b}{}^a = \g^a_{\a\b},\quad T_{\a\b}{}^\g = 0 ,}
where $\g^a$ are the symmetric gamma matrices in ten dimensions. In \BerkovitsUE. it was shown that \torscomp\ plus the Bianchi identities \bianchiC\ put the background to satisfy the background supergravity equations of motion.

\vskip 1cm

\noindent
{\bf Acknowledgments:} I would like to thank Nathan Berkovits, William Linch and Brenno Vallilo for useful comments. This work was partially financed by FONDECYT project 1120263.

\listrefs

\end